# Pathfinder first light: alignment, calibration, and commissioning of the LINC-NIRVANA ground-layer adaptive optics subsystem

Derek Kopon[*a], Al Conrad[a], Carmelo Arcidiacono[c], Tom Herbst[a], Valentina Viotto[b], Jacopo Farinato[b], Maria Bergomi[b], Roberto Ragazzoni[b], Luca Marafatto[b], Harald Baumeister[a], Thomas Bertram[a], Jürgen Berwein[a], Florian Briegel[a], Ralph Hofferbert[a], Frank Kittmann[a], Martin Kürster[a], Lars Mohr[a], Kalyan Radhakrishnan[a]

[a]Max-Planck-Institut für Astronomie, Königstuhl 17, D-69117 Heidelberg, Germany
[b]INAF, Osservatorio Astronomico di Padova, Vicolo Osservatorio 5, 35122 Padova, Italy
[c]INAF-Osservatorio Astronomico di Bologna, Via Ranzani 1, 40127 Bologna, Italy

## ABSTRACT

We present descriptions of the alignment and calibration tests of the Pathfinder, which achieved first light during our 2013 commissioning campaign at the LBT. The full LINC-NIRVANA instrument is a Fizeau interferometric imager with fringe tracking and 2-layer natural guide star multi-conjugate adaptive optics (MCAO) systems on each eye of the LBT. The MCAO correction for each side is achieved using a ground layer wavefront sensor that drives the LBT adaptive secondary mirror and a mid-high layer wavefront sensor that drives a Xinetics 349 actuator DM conjugated to an altitude of 7.1 km. When the LINC-NIRVANA MCAO system is commissioned, it will be one of only two such systems on an 8-meter telescope and the only such system in the northern hemisphere. In order to mitigate risk, we take a modular approach to commissioning by decoupling and testing the LINC-NIRVANA subsystems individually. The Pathfinder is the ground-layer wavefront sensor for the DX eye of the LBT. It uses 12 pyramid wavefront sensors to optically co-add light from natural guide stars in order to make four pupil images that sense ground layer turbulence. Pathfinder is now the first LINC-NIRVANA subsystem to be fully integrated with the telescope and commissioned on sky. Our 2013 commissioning campaign consisted of 7 runs at the LBT with the tasks of assembly, integration and communication with the LBT telescope control system, alignment to the telescope optical axis, off-sky closed loop AO calibration, and finally closed loop on-sky AO. We present the programmatics of this campaign, along with the novel designs of our alignment scheme and our off-sky calibration test, which lead to the Pathfinder's first on-sky closed loop images.

## 1. INTRODUCTION: LINC-NIRVANA, MCAO, AND THE PATHFINDER

LINC-NIRVANA[1,2] is the near-IR Fizeau interferometric beam combiner of the LBT being built by a consortium of German and Italian research institutions (Figure 1). The full LINC-NIRVANA will achieve 10 mas resolution (in J-band) over a 10 arcsec field of view using multi-conjugate adaptive optics (MCAO[3,4]) and fringe tracking[5]. This full system consists of multiple subsystems, each of which is complex in its own right: two ground-layer wavefront sensors[6,7] (GWS) consisting of 12 movable "star enlargers" (optical arms, each containing relay optics and a pyramid prism[8]); two mid-high layer wavefront sensors (MHWS) with eight star enlargers conjugated to ~7.1 km; and a cryostat containing the fringe tracker and IR science camera. These four wavefront sensors drive four deformable mirrors: the GWS's drive the adaptive secondary mirrors[9] and the mid-high wavefront sensors drive two Xinetics 349 actuator deformable mirrors. For reasons of risk mitigation and programmatic complexity, LINC-NIRVANA will be deployed in a modular fashion with the first subsystem, the Pathfinder[10,11], having begun commissioning in early 2013. Commissioning of the 2-layer MCAO system will commence in 2015, with the full interferometric capability planned for the future.

The Pathfinder (Figures 2&3) consists of the GWS[12,13] with 12 star enlargers that move independently to acquire natural guide stars, a blue steel support structure call "The Foot" that supports the GWS to the same height as the LINC-NIRVANA optical bench, an electronics cabinet, and an annular mirror that sends a 2-4 arcmin field to the GWS. The inner 2' field that passes through the annular mirror can be sent to an on-axis DSLR camera or an IR camera for acquisition and to measure Strehl.

---

[*] derek.kopon@gmail.com

As of this writing, our team has completed the first eight Pathfinder commissioning runs (T1-8), achieving single-star closed loop performance. The unfortunate weather conditions on Mt. Graham have hindered attempts at multiple star acquisition. However, full ground-layer multiple star correction is planned for the fall of 2014.

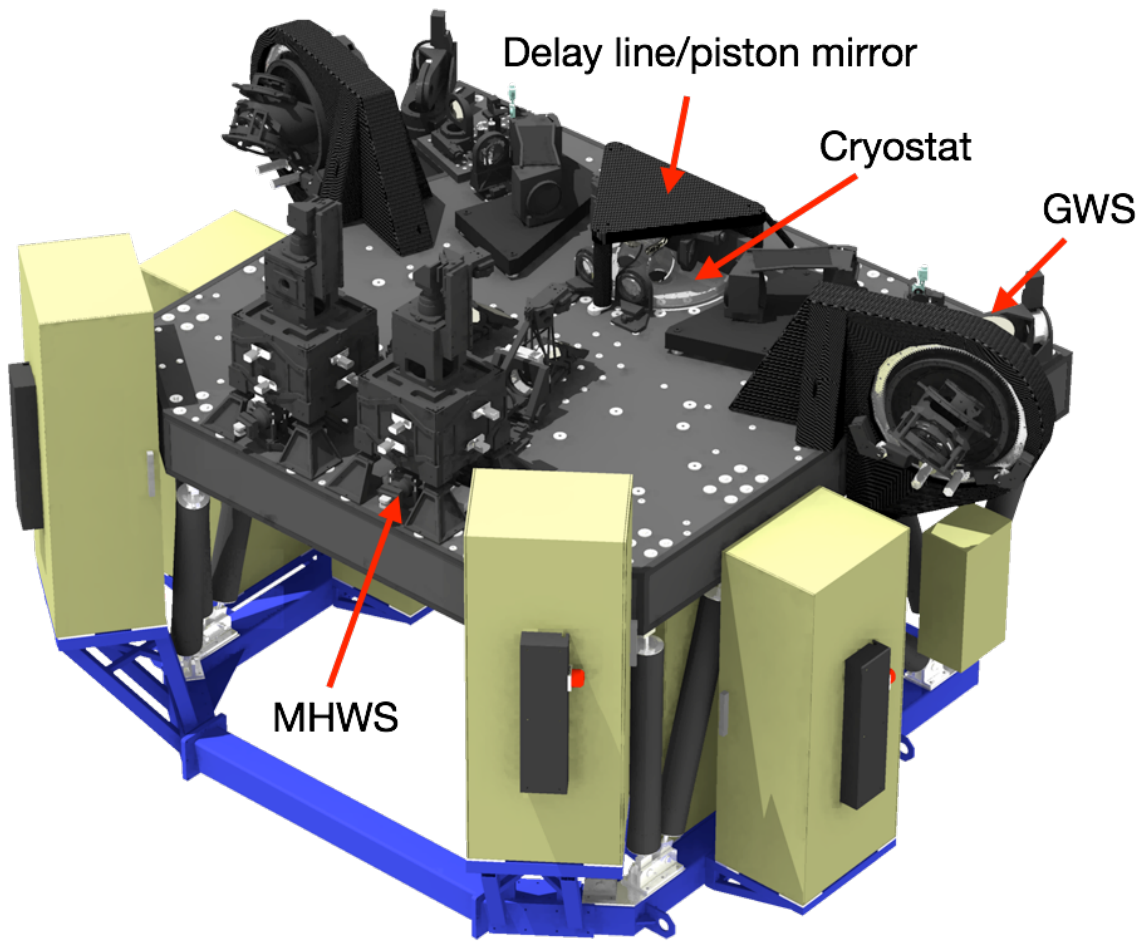

**Figure 1:** Model of the full LINC-NIRVANA instrument. The components on the "DX" side of the LBT are labeled and are mirror copies of those on the "SX" side. The full bench will be delivered to LBT in mid-2015. The DX GWS has already begun commissioning at the telescope as the Pathfinder experiment.

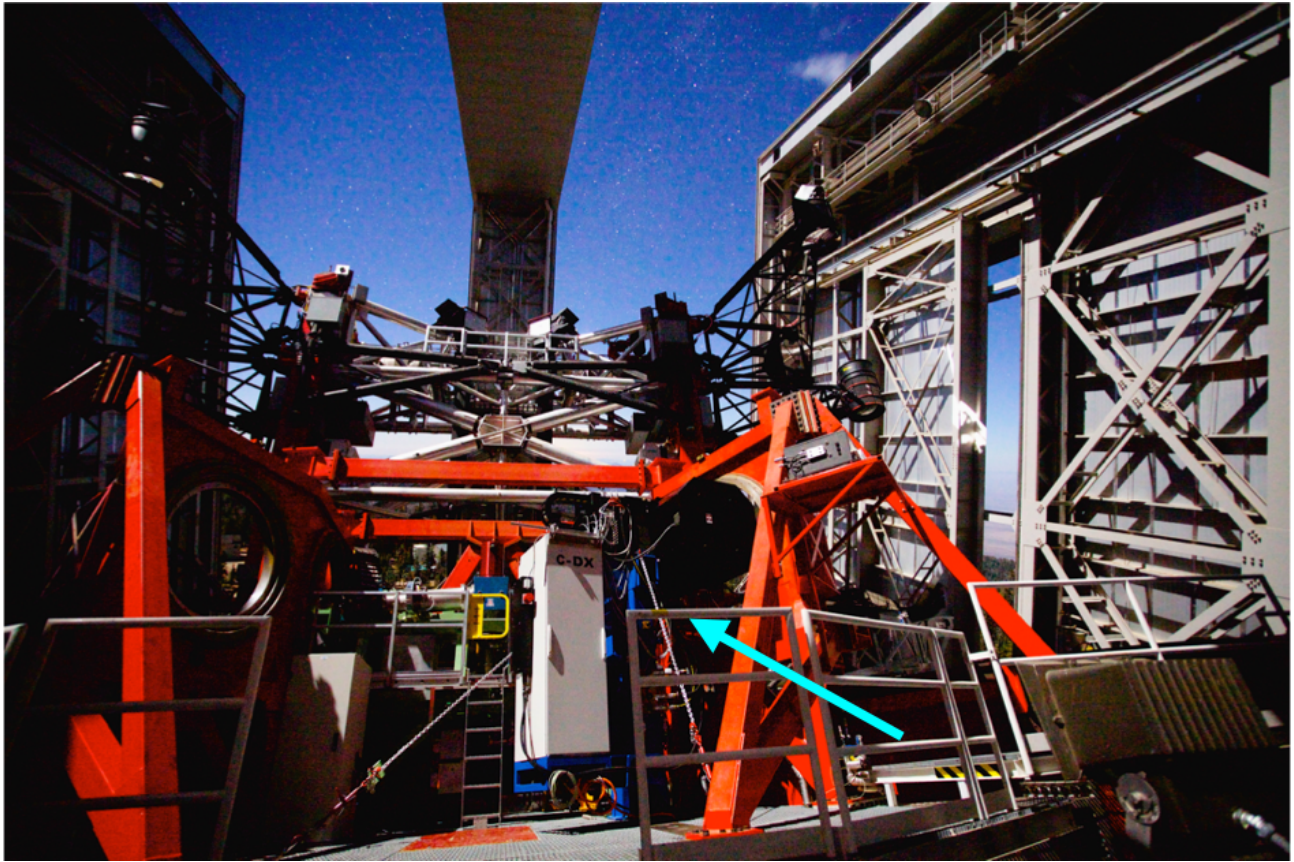

**Figure 2:** The Pathfinder mounted on the LBT "DX" side.

The 12 star enlargers move independently in a collision free manner to acquire bright guide star in the annular telescope focal plan. Each star enlarger consists of lenses and a pyramid that relay pupil images to a large common parabolic mirror. From there, a series of lenses reimage the pupil images to a common focal plane at the CCD50. In this manner, light from an arbitrary number of guide stars (up to 12) is optically coadded[14] to give one set of four pupils to allow wavefront sensing of the ground layer.

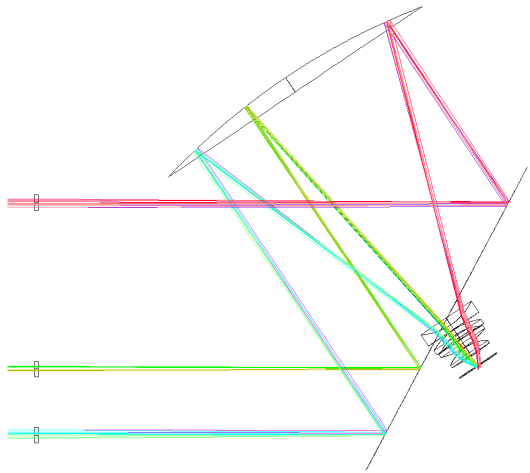 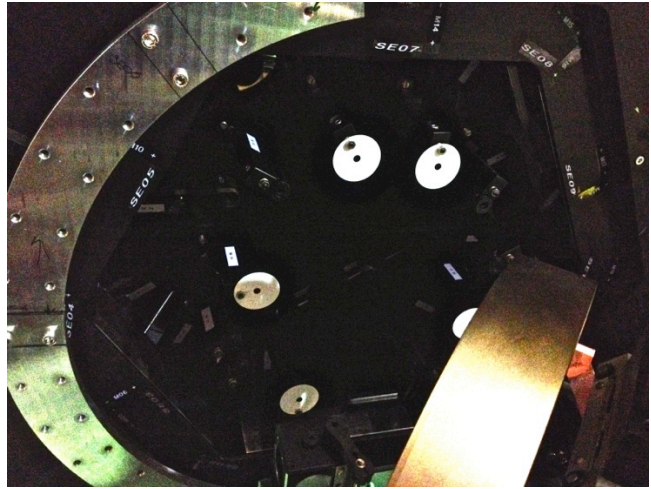

**Figure 3:** Optical path(s) of the Pathfinder. Left: Raytraces of three natural guide stars passing through three star enlargers and focused to a common focal plane. Right: Photograph of the star enlargers. The white "cheerio" disks are used in conjunction with a webcam for bright star acquisition during commissioning.

## 2. THE PATHFINDER 2013 AND 2014 COMMISSIONING CAMPAIGNS

The Pathfinder commissioning schedule is shown in Table 1. As of this writing, T8 was recently completed. However, poor weather again prevented multiple star acquisition. This will be the primary goal of the Fall 2014 runs.

| T1 | March 1-15, 2013 | Unpacking, assembly, and verification of rotator bearing accuracy and star enlarger accuracy. Magic lantern and annular mirror setup |
|---|---|---|
| T2 | April 3-11, 2013 | Star enlarger mapping and BCU software testing |
| T3 | April 24-28, 2013 | Craning of PF from the LBT mountain lab to telescope |
| T3.5 | June 17-22, 2013 | IRTC testing, fixturing, fit checking, and set up of alignment tools |
| T4 | Oct 1-8, 2013 | Alignment of PF mechanical rotation axis to the telescope optical axis |
| T5 | Nov 8-19, 2013 | Interaction matrix calibration and test with off-sky closed loop test. First closed loop image on a single star. |
| T6 | Dec 4-10, 2013 | Daytime: rotating interaction matrix, Night: attempt multiple star acquisition |
| T7 | March 29-April 4, 2014 | Daytime: rotating interaction matrix, Night: attempt multiple star acquisition |
| T8 | June 29-July 4, 2014 | Attempt multiple star acquisition[15,16], integrate IR camera |
| T9 | Fall 2014 | Characterize multiple star (GLAO) correction with IR camera |
| T10 | Fall 2014 | Characterize multiple star (GLAO) correction with IR camera |

**Table 1:** The Pathfinder 2013 and 2014 commissioning schedule. As of this writing, T8 was recently completed. However, multiple star acquisition was not achieved because of weather.

The early runs (T1-3) consisted mainly of unpacking, testing, and recalibrating the Pathfinder subsystems after shipment from Europe. Software interfaces with the LBT telescope control system were also tested, including slopes being sent from the GWS to the LBT adaptive secondary. For more detail on these tests, see Kopon et al. 2013.

## 3. PATHFINDER ALIGNMENT TO THE TELESCOPE

The alignment of the Pathfinder to the DX eye of the LBT created a unique challenge because of the lack of any on-axis surfaces or fiducials in the optical path of the GWS. The mechanical rotational axis of the GWS must be aligned to the optical axis of the telescope. To solve this problem, a rotating laser test was designed that involved mounting a laser on the GWS and reflecting it off the annular mirror, to the tertiary, and then to a translucent bead-blasted plastic screen either directly or after first reflecting off of the adaptive secondary (Figures 4&5). When the GWS bearing is rotated, the circle traced on the plastic screen by the laser coming directly from the tertiary designates a point in space through which the mechanical axis of the GWS bearing passes. When the laser is reflected off of the adaptive secondary, the circle traced by the laser on the screen designates a point through which the optical axis of the adaptive secondary (and therefore the telescope) passes (Figure 6). By adjusting the tip/tilt of the annular and tertiary mirrors, these two circles are brought to be concentric and the Pathfinder is thereby aligned to the telescope. This plan was realized during T4. During the rough alignment, binoculars are used to observe the circles traced by the laser (Figure 7). In the latter stages, an AVT camera on a post near the secondary is used to observe the plastic screen.

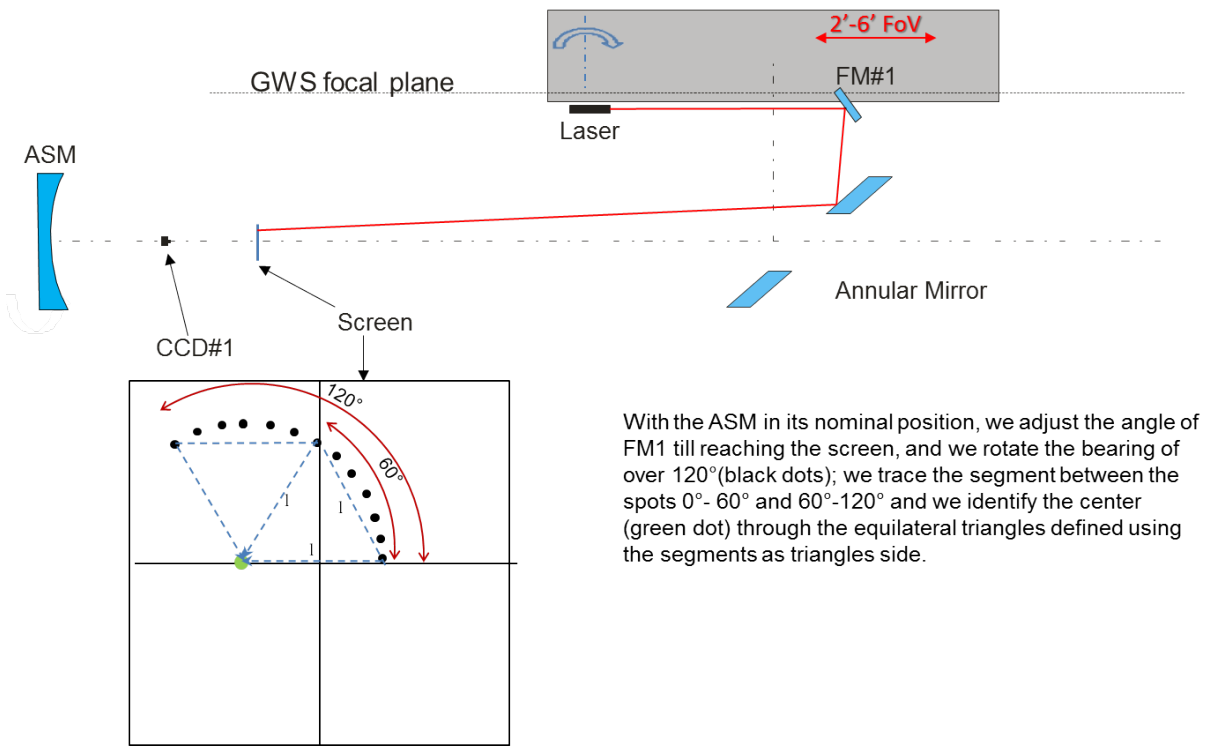

With the ASM in its nominal position, we adjust the angle of FM1 till reaching the screen, and we rotate the bearing of over 120°(black dots); we trace the segment between the spots 0°- 60° and 60°-120° and we identify the center (green dot) through the equilateral triangles defined using the segments as triangles side.

**Figure 4:** Configuration 1 defines the mechanical axis of the GWS rotator bearing by tracing a circle whose center is a point in space through which the axis passes.

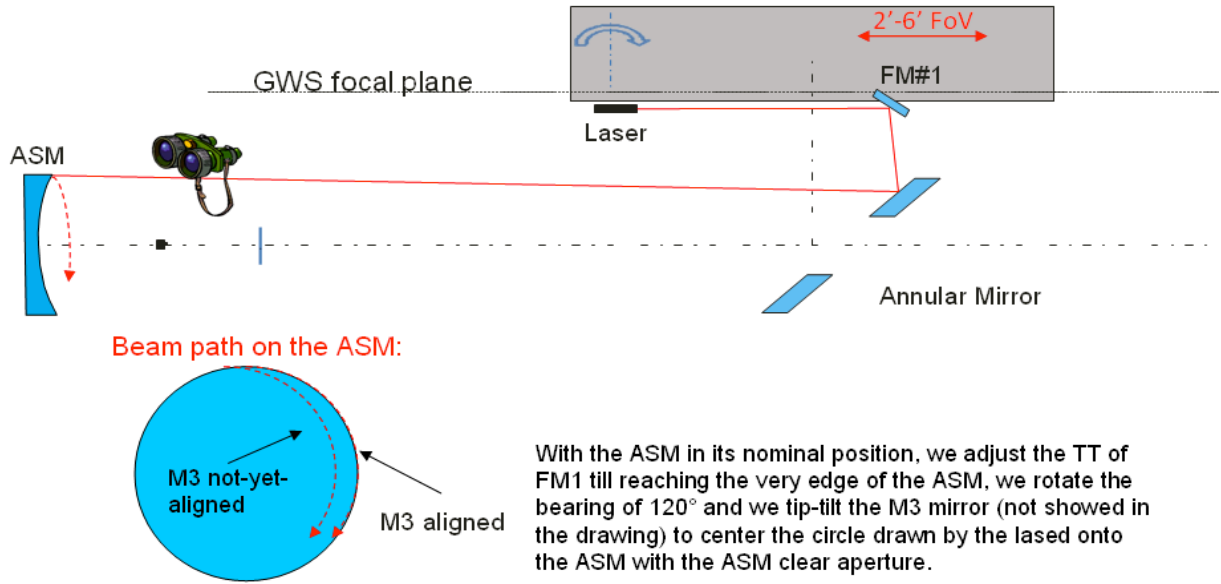

**Figure 5:** Configuration 2 defines the optical axis of the secondary mirror, and thereby the telescope, by tracing a circle whose center is a point in space through which the axis passes.

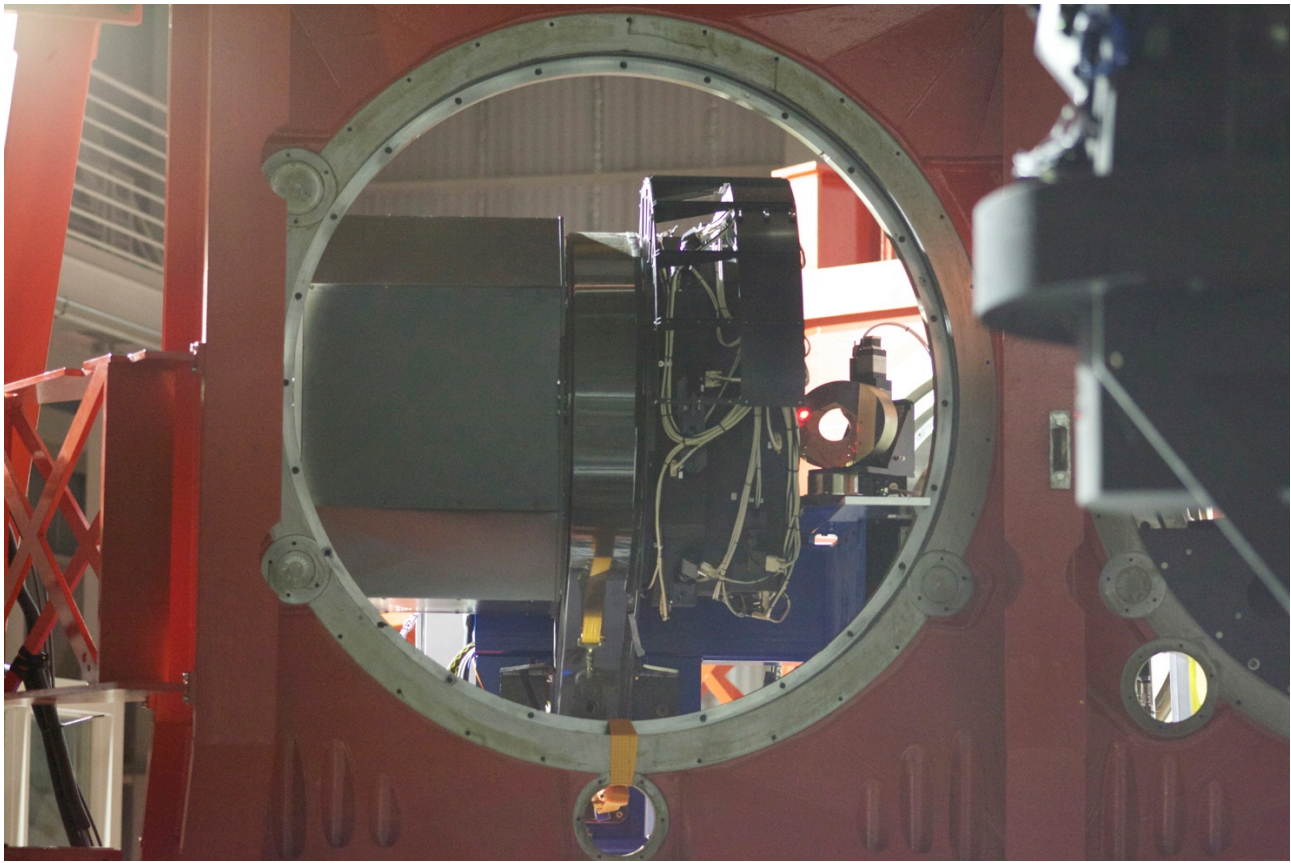

**Figure 6:** View of the Pathfinder from the far side of the DX primary. The red alignment laser can be seen reflecting off of the annular mirror before traveling to the tertiary.

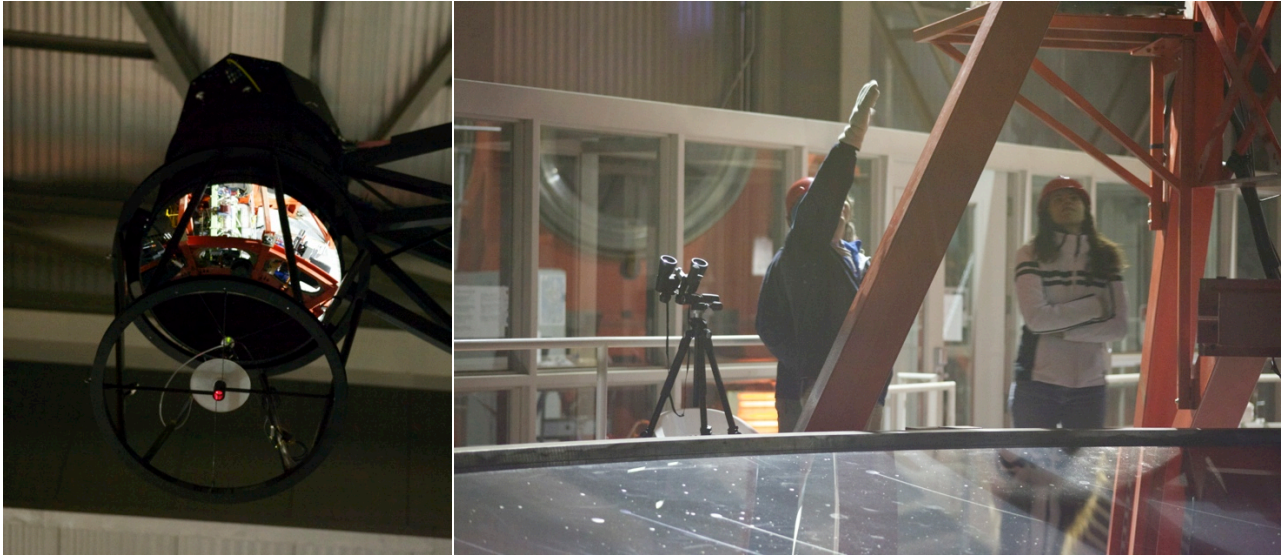

**Figure 7:** Left: The red alignment laser hitting the bead-blasted plastic screen located at the Gregorian intermediate focal plane. Right: Team members looking at the alignment circles with binoculars during the rough early stages of alignment.

## 4. DAYTIME/NIGHTTIME CLOSED LOOP TESTING

### 4.1 Daytime Double-Pass Closed-Loop Optical Test

We designed and implemented an off-sky closed loop test somewhat similar to that used by the LBT FLAO system[17]. This test is a double-pass optical configuration that begins with a multimode fiber located at the F/15 focal plane (Figure 8). The diameter of this fiber can be modified, but we typically use a 300 micron fiber diameter in order to create a spot at the wavefront sensor that is roughly seeing limited (0.5 arcsec). Light from the multimode fiber passes through the hole in the annular mirror, then through a beamsplitter, then to the tertiary, and then to the secondary where it is focused to a spot at the intermediate Gregorian focal plane and enters a retro-reflecting optic consisting of an on-axis parabolic mirror and a retro-reflecting flat (Figure 9). The light then travels back along the same path until reaching the beamsplitter, where it is reflected towards one of the star enlargers on the GWS. Because of the geometry of this test, in the simplest configuration, the illuminated star enlarger is located at the "9 o'clock" position in the GWS.

Initially, all of this alignment was performed with a bright (15 mW) green diode laser that could easily be seen by eye. After the return from the retroreflector was located near the F/15 focal plane, the tertiary mirror was adjusted to bring the portion of the return beam that was transmitted by the first beamsplitter back into the fiber. Next, gross coma was removed manually by tip/tilting and translating the secondary. Then, the tip/tilt of the beamsplitter was adjusted to send the reflected return beam into the travel range of the star enlarger. Once alignment had been achieved and laser light could be seen on the pupils on the CCD50, the green alignment laser was replaced with a white light source with a neutral density filter to avoid saturation and a long pass filter with a cut-on wavelength of 715 nm (Figure 10). A second beamsplitter just before the star enlarger allowed a small AVT camera to image the spot quality in a focal plane simultaneously as the pupils were illuminated.

The seeing-limited white light spot is then used to calculate the interaction matrix using a push/pull sequence. The AO loop can be tested by injecting turbulence into the adaptive secondary mirror. For more on the interaction matrix calculations of the AO system, see Bergomi et al. these proceedings.

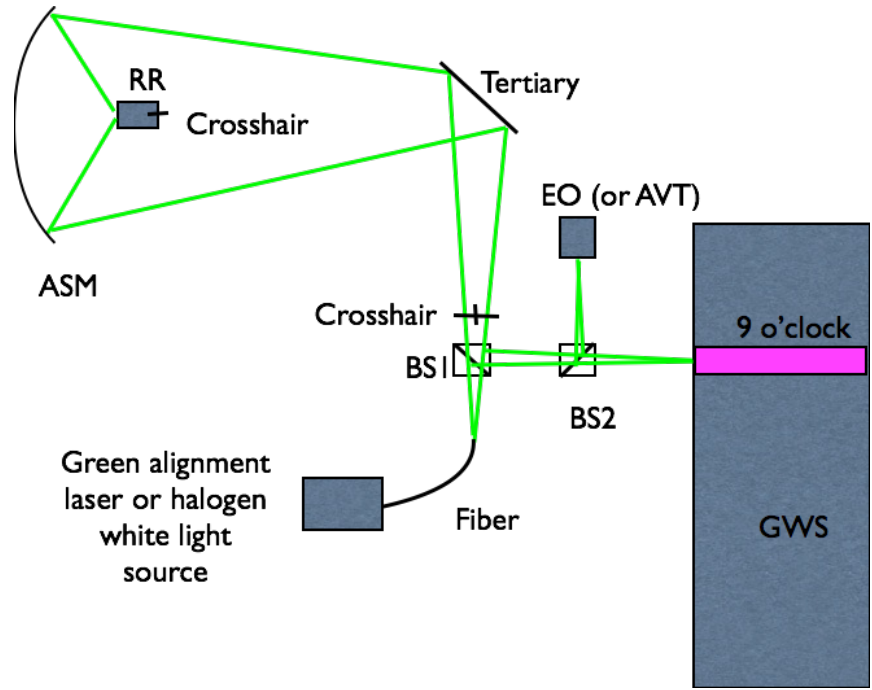

**Figure 8:** Double-pass optical test for running the AO system closed loop.

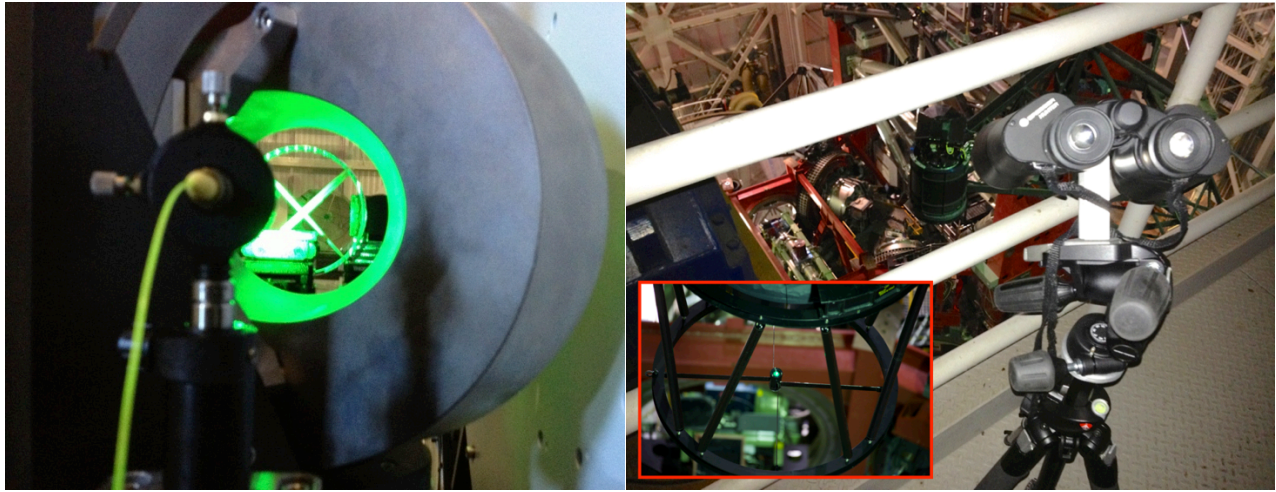

**Figure 9:** Left: View of the green alignment fiber laser passing through the hole in the annular mirror. Right: View from the 10[th] floor catwalk to guide the spot at the intermediate focal plane into the hole in the retro reflector.

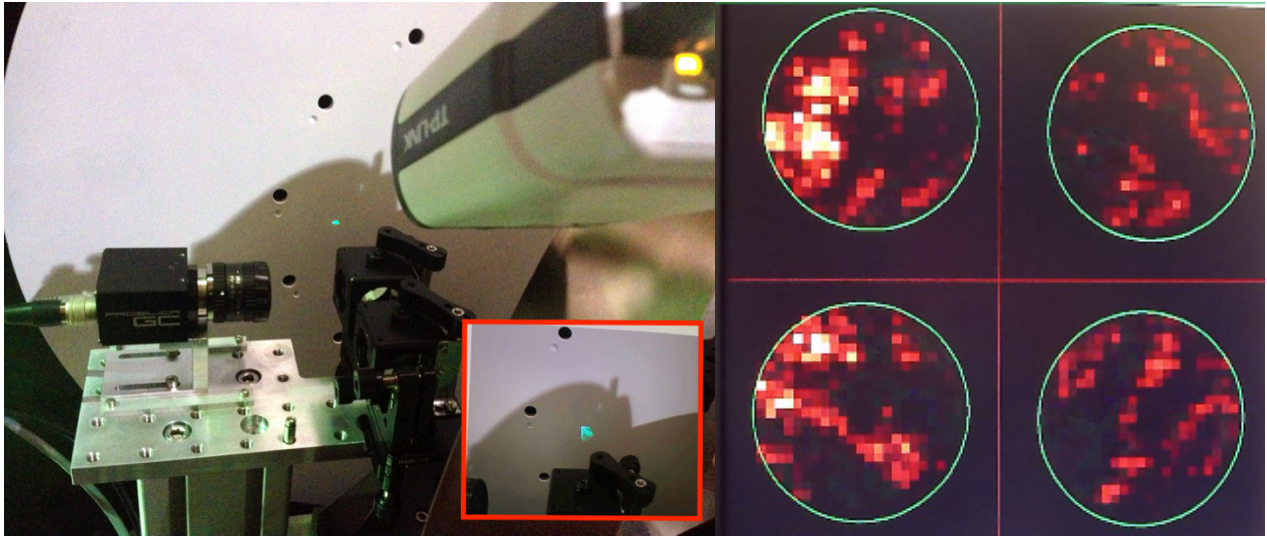

**Figure 10:** Left: A picture of the return spot from the alignment laser incident on the GWS cover. This is the spot prior to coma being removed with the positioning of the secondary mirror. Shown are the two beamsplitter cubes and the small camera that images the spot. Also shown in the top right foreground is a webcam that allows the spot on the cover to be viewed easily from the control room. Right: Simulated turbulence in the white light pupil images on the CCD50.

**4.2 Daytime Argos Single-Pass Optical Test**

A second daytime closed loop optical test that will be available to us in the future is the Argos calibration source[18]. This is a single-pass optical test that illuminates the secondary by placing fiber sources at the intermediate focal plane (see Figure 11). There is one on-axis source and three off-axis source located in our annular GWS field that are coma-corrected with computer generated holograms. The benefit of this test is largely logistical: the sources can be easily deployed on a swing arm with minimal time spent acquiring and aligning the source to the GWS. In contrast, the double pass test described in the previous section requires the installation of the retro reflector, the fiber, and the beamsplitter assembly, in addition to a few hours of acquisition and alignment. The Argos source is still in the process of being commissioned. Nonetheless, we were able to acquire an image and do some calibration with a spot from the on-axis fiber.

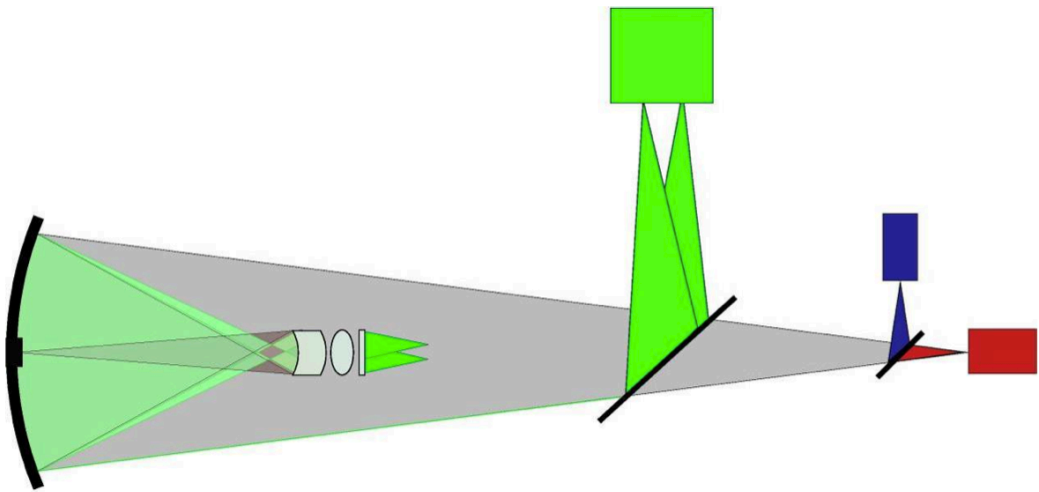

**Figure 11:** Schematic diagram of the Argos calibration source. The green box is a component used with the LUCI instrument configuration. For our purposes, the blue and red boxes represent our camera and WFS focal plane, respectively.

**4.3 Nighttime First Light Configuration**

Initially, we planned to obtain our first light images using an infrared test camera (IRTC). However, during the course of our T5 daytime tests, we realized that it would be mechanically and programmatically much simpler to slightly modify our daytime double pass setup for our first nighttime tests. By removing the "double-pass" components, namely the fiber and the retro reflector, we obtained a useful optical configuration with an on-axis Canon DSLR as our wide field acquisition camera and the pair of beamsplitters sending light simultaneously to the AVT camera and the GWS (see Figure 12). With this configuration, we obtained our first light images by closing the AO on a $3^{rd}$-magnitude star, Epsilon Aurigae (Figure 13). A 10-mode reconstructor was used initially, with the number of modes eventually increasing to 100.

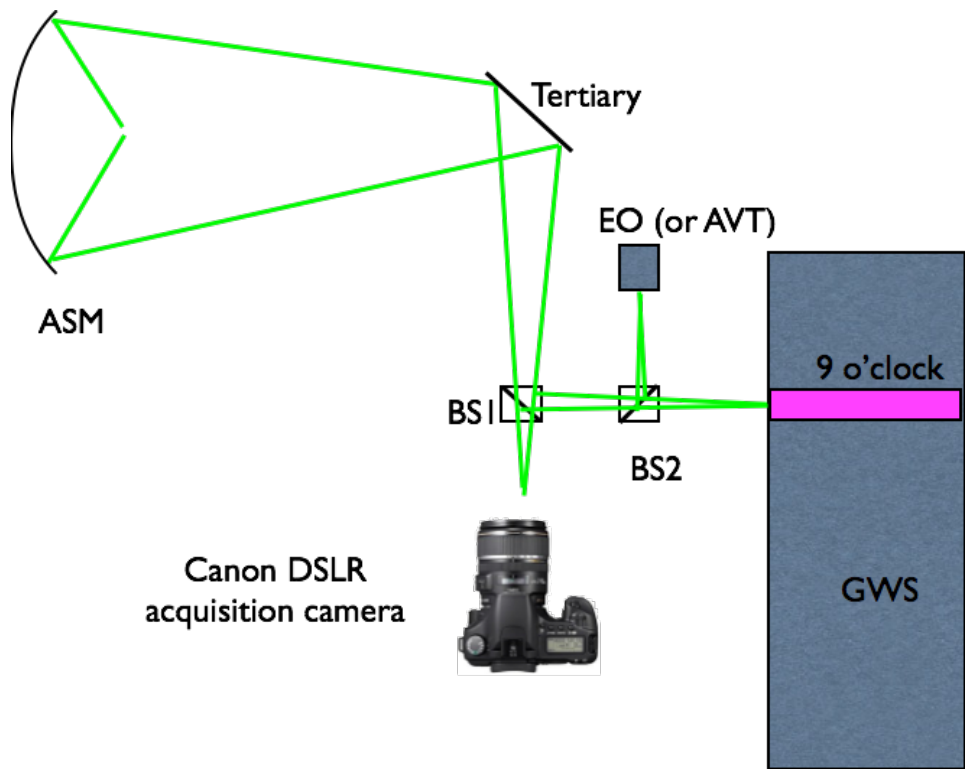

**Figure 12:** Optical configuration for the Pathfinder first light.

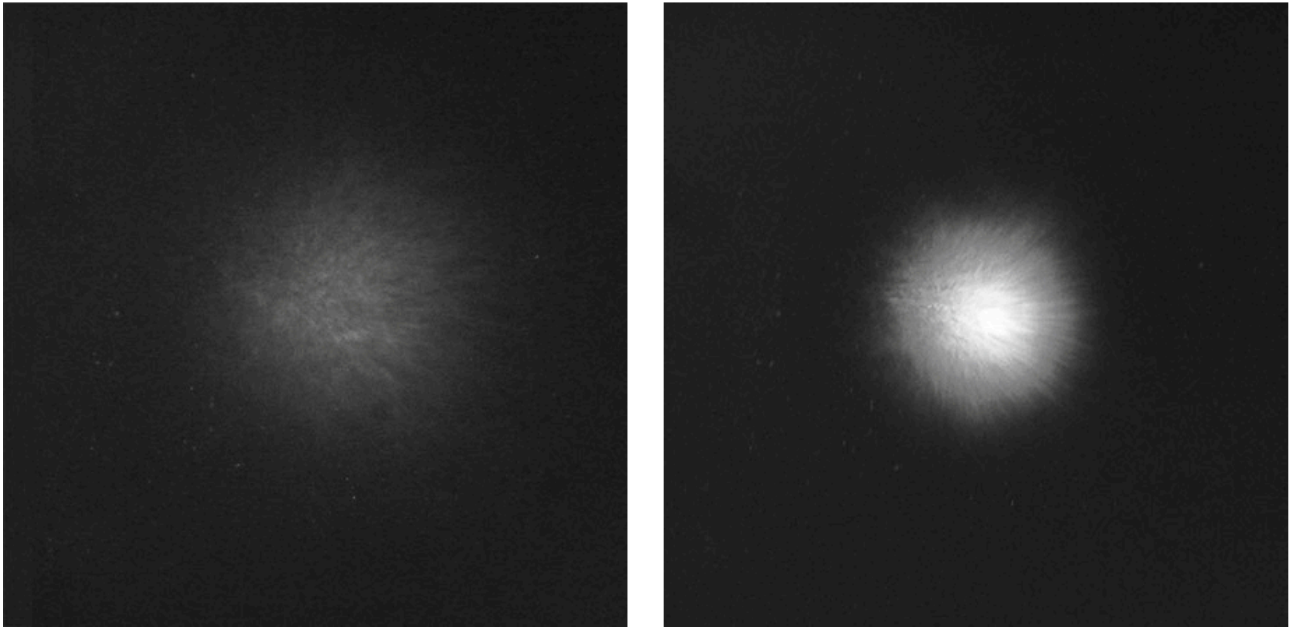

**Figure 13:** First light images of Epsilon Aurigae. Open loop image is shown on the left and closed loop on the right. The open loop seeing was 2.3 arcsec, quite poor.

## 5. CONCLUSION

The Pathfinder, the first LINC-NIRVANA AO subsystem at the LBT has achieved first light on a single guide star. Within the next few month, multiple star acquisition is planned in order to achieve ground layer correction over a 2 arcmin field. The full MCAO system will be deployed at the LBT in mid-2015.